\documentclass[preprint,showpacs,preprintnumbers,amsmath,amssymb]{article}

\usepackage{graphicx}
\usepackage{dcolumn}
\usepackage{bm}

\begin{document}

\title{Palatini Formalism of 5-Dimensional Kaluza-Klein Theory }

\author{You Ding \footnote{Email\ address:\ ding\_you@hotmail.com},
Yongge Ma \footnote{Email\ address:\ yonggema@yahoo.com}, Muxin
Han \footnote{Email\ address:\ hamsyncolor@hotmail.com}, and
Jianbing Shao \\
\small Department of Physics, Beijing Normal University, Beijing
100875, CHINA}

\date{\today}

\maketitle

\begin{abstract}
The Einstein field equations can be derived in $n$ dimensions
($n>2$) by the variations of the Palatini action. The Killing
reduction of 5-dimensional Palatini action is studied on the
assumption that pentads and Lorentz connections are preserved by
the Killing vector field. A Palatini formalism of 4-dimensional
action for gravity coupled to a vector field and a scalar field is
obtained, which gives exactly the same fields equations in
Kaluza-Klein theory.

\end{abstract}


{PACS number(s): 04.50.+h, 04.20.Fy}

\section{Introduction}
Palatini formalism is causing great interest in the study of
non-perturbative quantum gravity\cite{Ash}\cite{romano}\cite{AL},
modified gravity theories\cite{Fla}\cite{Vol} and their
cosmological applications\cite{MW}\cite{Kre}. Also, spacetime
reduction is very important in any high dimensional theory of
physics such as Kaluza-Klein
theory\cite{Bla}\cite{wesson}\cite{kk}\cite{Ma} and string
theory\cite{Polchinski} \cite{Yega}. The dimensional reduction can
make a high dimensional theory contact with the 4-dimensional
sensational world. The Campbell-Magaard theorem is generalized to
study the embedding of an $n$-dimensional spacetime into an
$(n+1)$-dimensional spacetime in high dimensional
physics\cite{seahra}. On the other hand, Killing reduction of
4-dimensional and 5-dimensional spacetimes have been studied by
Geroch\cite{Geroch} and Yang et. al.\cite{yang}. As in
Kaluza-Klein theory, the Killing reduction of 5-dimensional
Einstein spacetimes gives the 4-dimensional gravity coupled to the
electromagnetic field and a scalar field. The interesting scaler
field may contribute to the explanation to the dark physics in
current cosmology as well as the Higgs field in particle
physics\cite{krauss}\cite{wesson}.

Let ($M,g_{ab}$) be an $n$-dimensional spacetime with a Killing
vector field $\xi ^a$, which is everywhere spacelike. Let $S$
denote the collection of all trajectories of $\xi^a$. A map $\psi$
from $M$ to $S$ can be defined as follows: For each point $p$ of
$M$, $\psi(p)$ is the trajectory of $\xi^a$ passing through $p$.
Assume $S$ is given the structure of a differentiable
$(n-1)$-manifold such that $\psi$ is a smooth mapping. It is
natural to regard $S$ as a quotient space of $M$. The proof of
Geroch about the following conclusion is independent of the
dimension of $M$ \cite{yang}: There is a one-to-one correspondence
between tensor fields $\hat{T}_{a...c}^{b...d}$ on $S$ and tensor
fields $T_{a...c}^{b...d}$ on $M$ which satisfy
\begin{equation}
\begin{array}{l}
\xi^a T^{b\cdots d}_{a\cdots c}=0,\ \cdots \ ,\xi_d T^{b\cdots
d}_{a\cdots
  c}=0 ,\\
{\mathcal{L}}_\xi T^{b\cdots d}_{a\cdots c}=0 , \label{Li D}
\end{array}
\end{equation}where ${\mathcal{L}}_\xi$ denotes the Lie derivative
with respect to $\xi^a$. The entire tensor field algebra on $S$ is
completely and uniquely mirrored by tensor field on $M$ subject to
Eq. (\ref{Li D}). Thus, we shall speak of tensor fields being on
$S$ merely as a shorthand way of saying that the fields on $M$
satisfy Eq. (\ref{Li D}). The metric and the Kronecker delta on
$S$ are defined as
\begin{eqnarray}
&&h_{ab}=g_{ab}-\lambda ^{-1}\xi _a\xi_b,\label{h_{ab}}\\[2pt]
&&h_a^b=\delta _a^b-\lambda ^{-1}\xi _a\xi ^b, \label{h_a^b}
\end{eqnarray}
where $\lambda \equiv\xi ^a\xi _a$. Eq. (\ref{h_a^b}) can also be
interpreted as the projection operator onto $S$. Note that in
general $S$ cannot be an embedded submanifold of $M$\cite{yang},
hence the Campbell theorem is not valid for this treatment. Note
also that the abstract index notation\cite{wald}\cite{Liang} is
employed for spacetime indices through this paper.

To study the Palatini formalism of 5-dimensional Kaluza-Klein
theory, we first extremize the $n$-dimensional Palatini action and
obtain the pure Einstein field equations. Then, we reduce the
5-dimensional Palatini action, assuming there is a spacelike
Killing vector field in the 5-dimensional spacetime. Note that if
the extra dimension is compactified as a circle $S^1$ with a
microscopic radius, a Killing vector field may arise naturally in
low energy regime\cite{Bla}. Since we are working in Palatini
formalism, besides the assumption that the connection and pentad
are preserved by the Killing vector field, we also have to assume
certain relation of the 4-dimensional electromagnetic field as
well as scalar field and some components of the underlying
5-dimensional connection. This is motivated by the pentad
formalism. By the Killing reduction, we obtain a Palatini
formalism of 4-dimensional action coupled to a vector field and a
scalar field. The variations of this action give the coupled
fields equation, which are as same as those in the 5-dimensional
Kaluza-Klein theory.

\section{Palatini action in $n$ dimensions}

In this section, following the approach in 4
dimensions\cite{Ash}\cite{romano}, we will show in detail  that
the Palatini action reproduce Einstein's equation in $n$
dimensions ($n>2$). Although this is a well-known result, to our
knowledge the same proof for $n$ dimensions has not appearred so
far in the literature.

  Consider an $n$-manifold $M$, on which the basic dynamical
variables in the Palatini framework are $n$-bases $(e_{\mu})^{a}$
and Lorentz connections $\omega_{\ \ a}^{\mu\nu}$, where the Greek
indices $\mu,\nu$ denote the internal Lorentz group. The internal
space is equipped with a Minkowskian metric $\eta_{\mu\nu}$ (of
signature $- + \ldots +$),  which is fixed once and for all.
Consequently, one can freely raise and lower the internal indices;
their position does not depend on the choice of dynamical
variables. To raise or lower the spacetime indices $a,b,\ldots$,
on the other hand, one needs a space-time metric $g_{ab}$ which is
a dynamical variable, constructed from the duel bases
$(e^{\mu})_{a}$ via:
\begin{displaymath}
g_{ab}=\eta_{\mu\nu}(e^{\mu})_{a}(e^{\nu})_{b}.
\end{displaymath}
The connection 1-form $\omega_{\ \ a}^{\mu\nu}$ acts only on
internal indices; it defines a generalized derivative operator
$\tilde{\nabla}_a$ via:
\begin{equation}
\tilde{\nabla}_{a}K_{\mu}:=\partial_{a}K_{\mu}+\omega_{\mu\ a}^{\
\ \nu}K_{\nu},\label{deriv}
\end{equation} where $\partial_{a}$ is a fiducial derivative
operator. Since $\tilde{\nabla}_a$ annihilates the fiducial
Minkowskian metric $\eta_{\mu\nu}$ on the internal space, the
connection 1-forms $\omega_{\ \ a}^{\mu\nu}$ are antisymmetric in
$\mu$ and $\nu$; they take values in the Lorentz Lie algebra. The
curvature $\Omega_{ab}^{\ \ \mu\nu}$ of the connection $\omega_{\
\ a}^{\mu\nu}$ is given by:
\begin{displaymath}
\Omega_{ab}^{\ \ \mu\nu} =
(d\omega^{\mu\nu})_{ab}+[\omega_{a},\omega_{b}]^{\mu\nu},
\end{displaymath}
where $[,]$ stands for the commutator in the Lorentz Lie algebra.

  The Palatini action is given by:
\begin{equation}
S_{p}[(e_{\mu})^{a},\omega_{\ \ a}^{\mu\nu}] =
\int_{M}e(e_{\mu})^{a}(e_{\nu})^{b}\Omega_{ab}^{\ \ \mu\nu}
\label{action},
\end{equation}
where $e$ is the square root of the determinant of the $n$-metric
$g_{ab}$. The field equations are obtained by varying this action
with respect to $(e_{\mu})^{a}$ and $\omega_{\ \ a}^{\mu\nu}$. To
carry out the variation with respect to the connection, it is
convenient to introduce the unique (torsion free) connection
$\nabla_{a}$ on both space-time and internal indices determined by
the bases $(e_{\mu})^{a}$ via:
\begin{equation}
\nabla_{a}(e_{\mu})^{b}=0 \label{basis} .
 \end{equation}
  The
difference between the actions of $\nabla_{a}$ and
$\tilde{\nabla}_a$ on internal indices is characterized by a field
$C_{a\mu}^{\ \ \nu}$:
\begin{equation}
(\tilde{\nabla}_{a}-\nabla_{a})V_{\mu}=C_{a\mu}^{\ \ \nu}V_{\nu}.
\label{C1}
\end{equation}
The difference between their curvatures is given by:
\begin{equation}
\Omega_{ab}^{\ \ \mu\nu}-R_{ab}^{\ \ \mu\nu}
=2\nabla_{[a}C_{b]}^{\ \mu\nu}+2C_{[a}^{\ \mu\rho}C_{b]\rho}^{\ \
\nu},\label{C2}
\end{equation}
where $R_{ab}^{\ \ \mu\nu}$ is the internal curvature of
$\nabla_{a}$. Note that the variation of the action with respect
to $\omega_{\ \ a}^{\mu\nu}$ (keeping the basis fixed) is the same
as the variation with respect to $C_{a}^{\ \mu\nu}$. Using Eq.
(\ref{C2}), the Palatini action (\ref{action}) becomes:
\begin{equation}
S_{p}[(e_{\mu})^a,C_a^{\ \mu\nu}] =
\int_{M}e(e_{\mu})^{a}(e_{\nu})^{b} (R_{ab}^{\ \ \mu\nu} +
2\nabla_{[a}C_{b]}^{\ \mu\nu}+2C_{[a}^{\ \mu\rho}C_{b]\rho}^{\ \
\nu}).
\end{equation}
By varying this action with respect to $C_a^{\ \mu\nu}$, one
obtains:
\begin{equation}
\big((e_{\rho})^{[a}(e_{\sigma})^{b]} \delta^{\rho}_{[\mu}
\delta_{\nu]}^{\tau}\big)C_{b\tau}^{\ \ \sigma}=0,\label{last}
\end{equation}
We now show that Eq. (\ref{last}) implies:
\begin{equation}
C_{a\mu}^{\ \ \nu}\,=\,0\label{C0}. \end{equation}

 To see this, define a space-time tensor field
$S_{abc}:= C_{a\mu\nu}(e^{\mu})_{b}(e^{\nu})_{c}$. Then the
condition $C_{a\mu\nu}=C_{a[\mu\nu]}$ is equivalent to
$S_{abc}=S_{a[bc]}$. Now contracting Eq. (\ref{last}) with
$(e^{\mu})_{a}(e^{\nu})_{c}$, one obtains
\begin{equation}
(n-2)S_{bc}^{\ \ b}+S_{ca}^{\ \ a}=0
\end{equation}
 This yields $S_{bc}^{\ \ b}=0$, when $n\neq2$. Hence $S_{abc}$ is
trace-free on its first and last indices. Using this result, Eq.
(\ref{last}) leads to
\begin{equation}
C_{b\mu}^{\ \ \rho}(e_{\rho})^{a} (e_{\nu})^{b} - C_{b\nu}^{\ \
\rho}(e_{\mu})^{b}(e_{\rho})^{a}=0 . \label{last2}
\end{equation}
If we now contract Eq. (\ref{last2}) with
$(e^{\mu})_{c}(e^{\nu})_{d}$, we get
\begin{equation}
S_{cd}^{\ \ a}=S_{(cd)}^{\ \ \ a}
\end{equation}
Thus, $S_{abc}$ is symmetric in its first two indices. Since
$S_{abc}=S_{a[bc]}$ and \\ $S_{abc}=S_{(ab)c}$, we can
successively interchange the first two indices to show
$S_{abc}=0$. Furthermore, since $(e^{\mu})_{a}$ are invertible, we
get $C_{a\mu}^{\ \ \nu}=0$. This is the desired result.

 Thus, the equation of motion for the derivative operator
$\tilde{\nabla}_{a}$ is simply that it equals $\nabla_{a}$ while
acting on objects with only internal indices. Thus the connection
$\tilde{\nabla}_{a}$ is completely determined by the bases.  By
carrying out the variation of action (\ref{action}) with respect
to the bases, one obtains:
\begin{equation}
(e_\mu)^c\,\Omega_{cb}^{\ \ \mu\nu}\,-\,\frac12\,\Omega_{cd}^{\ \
\rho\sigma}\,(e_\rho)^c\,(e_\sigma)^d(e^\nu)_b\,=\,0
\label{Einstein}
\end{equation} Substitution of Eq. (\ref{C0}) in
Eq. (\ref{C2}) implies that $\Omega_{ab}^{\ \ \mu\nu}=R_{ab}^{\ \
\mu\nu}$. Using the fact that the internal curvature of $\nabla_a$
is related to its space-time curvature by $R_{ab\mu}^{\ \ \
\nu}=R_{abc}^{\ \ \ d}\,(e_{\mu})^{c}\,(e^{\nu})_{d}$ and
multiplying Eq. (\ref{Einstein}) by $(e_{\nu})_a$ tells us that
the Einstein tensor $G_{ab}$ of the metric $g^{ab}$ vanishes.

\section{Palatini action in 5-dimensional Kaluza-Klein theory}

To make Killing reduction of the Palatini action (\ref{action}),
we first generalize the reduction program in Refs.\cite{Geroch}
and \cite{yang} to generalized tensor fields.

Suppose $S$ is the reduced manifold of $n$-dimensional spacetime
($M,g_{ab}$) with a spacelike Killing vector field $\xi^a$ as in
Sec.1. Let $(e_{\mu})^a$ ($\mu = 0,1,\cdots,n-1$) be orthonormal
bases on $M$. To simplify the formalism, we make a partial gauge
fixing by choosing
\begin{equation}
(e_{n-1})^a=\lambda^{-\frac12}\xi^a,\label{en-1}
\end{equation}and assume
\begin{equation}
{\mathcal{L}}_\xi \ {\omega}^{\mu\nu}_{\ \ a}=0,\
{\mathcal{L}}_\xi ({e}_{i})^a=0,\ i=0,1,\cdots,n-2.\label{Lie}
\end{equation}
It is easy to see that Eq. (\ref{en-1}) implies ${\mathcal{L}}_\xi
({e}_{n-1})^a=0$ and $\xi_{a}({e}_{i})^a=0$. Eq. (\ref{Lie}) then
means that $(e_i)^a$ are orthonormal bases on $S$, denoted by
$(\hat{e}_i)^a$. It is natural to think $i,j,k=0,1,\cdots,n-2$ as
internal Lorentz indices on $S$. Using Eq. (\ref{Lie}), one can
define Lorentz connection 1-forms $\hat{\omega}^{ij}_{\ \ a}$ on
$S$ as
\begin{equation}
\hat{\omega}^{ij}_{\ \ a} = h^b_a \omega^{ij}_{\ \
b}.\label{homega}
\end{equation}It is easy to see that there is a one-to-one
correspondence between generalized tensor fields $\hat{T}^{b\cdots
d i \cdots j}_{a\cdots c k \cdots l}$ on $S$ and generalized
tensor fields $T^{b\cdots d i \cdots j}_{a\cdots c k \cdots l}$ on
$M$ which satisfy
\begin{equation}
\begin{array}{l}
\xi^a T^{b\cdots d i \cdots j}_{a\cdots c k \cdots l}=0,\ \cdots \
,\xi_d T^{b\cdots d i \cdots j}_{a\cdots c k \cdots l}=0 ,\\
{\mathcal{L}}_\xi T^{b\cdots d i \cdots j}_{a\cdots c k \cdots
l}=0. \label{LiD}
\end{array}
\end{equation}

A generalized derivative on $S$ is defined by
\begin{displaymath}
\tilde{D}_aT_{c_1\cdots c_lj_1\cdots j_n}^{b_1\cdots b_ki_1\cdots
i_m}=h_{a}^{a_1}h_{d_1}^{b_1}\cdots
h_{d_k}^{b_k}h_{c_1}^{e_1}\cdots h_{c_l}^{e_l}\tilde{\nabla}
_{a_1}T_{e_1\cdots e_l j_1\cdots j_n}^{d_1\cdots d_ki_1\cdots
i_m},
\end{displaymath}
where $\tilde{\nabla} _{a}$ is the generalized  derivative on $M$
defined by Eq. (\ref{deriv}) and $T_{c_1\cdots c_lj_1\cdots
j_n}^{b_1\cdots b_ki_1\cdots i_m}$ is any generalized tensor field
on $S$. Note that $\tilde{D}_a$ satisfies all the conditions of a
derivative operator and
\begin{displaymath}
\tilde{D}_aV_i=\,\hat{\partial}_aV_i+\hat{\omega}^{\ j}_ {i\
a}V_j,
\end{displaymath}
where $\hat{\partial}_a$ is the fiducial derivative operator on
$S$ defined by $\partial_a$ on $M$. The unique connection
determined by $(\hat{e}_i)^a$ on $S$ reads
\begin{displaymath}
{D_a}V_i = h^b_a{\nabla}_bV_i,
\end{displaymath}
where ${\nabla}_a$ is the connection on $M$
defined by Eq. (\ref{basis}).

We now consider the special case where $n$=5. Let $\varepsilon
_{abcde}$ be the volume element associated with the metric
$g_{ab}$
on $M$. Then it can be shown that \\
$\varepsilon _{abcd}\equiv |\lambda|^{-\frac{1}{2} }\varepsilon
_{abcde}\xi ^e$ is the volume element associated with the metric
$h_{ab}$ on $S$\cite{yang}. Let
\begin{equation}
F_{ab}\equiv -\frac 12 \lambda ^{-\frac 32}\varepsilon
_{abcd}\omega^{cd},\label{Fab}
\end{equation}
where the twist 2-form of $\xi ^a$ is defined as $
\omega_{ab}:=\varepsilon _{abcde}\xi^c\nabla ^d\xi^e$. Clearly we
have $F_{ab}=F_{[ab]}$ and $F_{ab}\xi ^a=0$. It is also easy to
see that
\begin{equation}
\mathcal{L} _\xi \lambda =0,\hspace{1cm} \mathcal{L} _\xi
F_{ab}=0. \label{Li lambda}
\end{equation}
Hence, $\lambda $ and $F_{ab}$ are fields on $S$. It is shown in
Ref.\cite{yang} that there is at least locally a one-form $A_a$ on
$S$ such that $F_{ab}=(dA)_{ab}$, which will be shown to play
still the role of electromagnetic field on $S$.

Suppose the Lorentze connection ${\omega}^\mu_{\ \nu a}$ be
compatible with the pentad $(e^\nu)_b$. It is then easy to see
that the reduced connection $\hat{\omega}^i_{\ ja}$ on $S$ would
be compatible with the tetrad $(\hat{e}^j)_b$. One thus has the
structure equations\cite{kk}
\begin{equation}
(d{e}^\mu)_{ab} = -{\omega}^\mu_{\ \nu a}\ \wedge\
(e^\nu)_b\label{demu}
\end{equation}
and
\begin{equation}
(d{\hat{e}^i})_{ab} = -\hat{\omega}^i_{\ ja}\ \wedge\
(\hat{e}^j)_b\label{dei}.
\end{equation}
It is easy to check $\xi^a(de^\mu)_{ab}=0$ , which leads to
\begin{equation}
 (d{e}^i)_{ab}\equiv-\omega^i_{\ ja}\wedge(e^j)_b-\omega^i_{\ 4a}\wedge(e^4)_b=(d\hat{e}^i)_{ab}.\label{equal}
\end{equation}
From Eq. (\ref{homega}), one has
\begin{equation}
\hat{\omega}^{ij}_{\ \ a}(\hat{e}_k)^a\equiv\hat{\omega}^{ij}_{\ \
k}=\omega^{ij}_{\ \ k}\equiv {\omega}^{ij}_{\ \ a}
({e}_k)^a.\label{coeffient}
\end{equation}
Substituting Eqs. (\ref{dei}) and (\ref{coeffient}) into Eq.
(\ref{equal}), one gets
\begin{displaymath}
\omega^i_{\ 4a}\wedge(e^4)_b=0,
\end{displaymath}which leads to
\begin{equation}
\omega^i_{\ [4j]}=0.\label{wi[4j]}
\end{equation}
Using Eq. (\ref{wi[4j]}), one obtains
\begin{equation}
\omega_{4ij}=-\omega_{ij4}, \hspace{1cm} \omega^4_{\
ij}=\omega^4_{\ [ij]}\label{omega4}.
\end{equation}
The exterior derivative of the remaining basis is given by
\begin{equation}
d({e}^4)_{ab} = -{\omega}^4_{\ ia}\wedge ({e}^i)_b.\label{de4}
\end{equation}
Using Eqs. (\ref{en-1}) and (\ref{Fab}), it can also be expressed
as
\begin{equation}
(de^4)_{ab}
=2\nabla_{[a}\lambda^{-\frac12}\xi_{b]}=\lambda^{-\frac
32}\xi_{[b}D_{a]}\lambda+\lambda^{\frac12}F_{ab}.\label{de4 }
\end{equation}
 By using Eqs. (\ref{omega4}), (\ref{de4}) and (\ref{de4 }), one obtains
\begin{eqnarray}
-\omega_{ij4}=\omega^4_{\ ij}=\frac 12\lambda^{\frac12}F_{ij}, \label{w4[ij]}\\
\omega^4_{\ i4}=\frac 12\lambda^{-1}D_i\lambda, \label{w4i4}
\end{eqnarray}
where $D_i\lambda\equiv(e_i)^aD_a\lambda,\ F_{ij}\equiv
F_{ab}(e_i)^a(e_j)^b$.

Although ${\omega}^\mu_{\ \nu a}$ is not necessarily compatible
with $(e^\nu)_b$ in the Palatini formalism, we will still take
Eqs. (\ref{w4[ij]}) and (\ref{w4i4}) as an assumption. Using Eqs.
(\ref{coeffient}) and (\ref{w4[ij]}), one thus gets the
relationship
\begin{equation}
{\omega}^i_{\ ja}=\hat{\omega}^i_{\ ja}+\frac12
\lambda^{\frac12}F_{j}^{\ i}({e}^4)_a\label{wij}
\end{equation}
between the connections in $M$ and $S$. And using Eqs.(\ref{w4i4})
and (\ref{w4[ij]}), one obtains
\begin{equation}
{\omega}^4_{\ ia}=\frac12 \lambda^{\frac12}F_{ij}({e}^j)_a +
\frac12\lambda^{-1} D_i\lambda ({e}^4)_a.\label{w4i}
\end{equation}

The curvature 2-forms of connections $\omega^{\mu\nu}_{\ \ a}$ on
$M$ are defined by the structure equation
\begin{displaymath}
{\Omega}_{ab}^{\ \ \mu\nu}= (d{\omega}^{\mu\nu})_{ab}+
{\omega}^{\mu}_{\ \lambda a}\ \wedge\ {\omega}^{\lambda \nu}_{\ \
b} \equiv \frac12\ {\Omega}^{\ \ \mu\nu}_{\rho\sigma} \
(e^{\rho})_a\wedge\ ({e}^\sigma)_b.
\end{displaymath}
Using Eqs. (\ref{wij}) and (\ref{w4i}), one gets
\begin{eqnarray}
{\Omega}_{ab}^{\ \ i4}&=& (d{\omega}^{i4})_{ab}+ {\omega}^{i}_{\ j
a}\ \wedge\ {\omega}^{j 4}_{\ \ b} \label{wabi4}\\ \nonumber
&=&\frac12\big(\frac12 \lambda^{\frac12}\tilde{D}^iF_{lk}+\frac14
\lambda^{-\frac12}(2F_{lk}D^i\lambda+F_l^{\ i}D_k\lambda-F_k^{\
i}D_l\lambda)\big)(e^k)_a\wedge(e^l)_b \\ \nonumber &&+
\frac12\big(\frac14 \lambda F^{ij}F_{kj}-
\frac12\lambda^{-\frac12}\tilde{D}_k(D^i\lambda)+
\frac14{\lambda}^{-2}(D^i\lambda)(D_k\lambda)\big)(e^k)_a\wedge(e^4)_b
\end{eqnarray}
and
\begin{eqnarray}
{\Omega}_{ab}^{\ \ ij}&=& (d{\omega}^{ij})_{ab}+ {\omega}^{i}_{\ k
a}\ \wedge\ {\omega}^{k j}_{\ \ b}+\omega^{i}_{\ 4a}\wedge\omega^{4j}_{\ \ b} \label{wabij}\\
\nonumber &=&\hat{\Omega}^{\ \ ij}_{ab} +\frac12\big(
\frac14\lambda(2F^{\ ji}F_{kl}-F_k^{\ i}F_{l}^{\ j}-F_l^{\ i}F_{\
k}^j)\big)(e^k)_a\wedge(e^l)_b\\ \nonumber &&+\frac12\big(\frac12
\lambda^{\frac12}\tilde{D}_kF^{ji}+\frac14
\lambda^{-\frac12}(2F^{ji}D_k\lambda+F^j_{\ k}D^i\lambda-F^i_{\
k}D^j\lambda)\big)(e^k)_a\wedge(e^4)_b.
\end{eqnarray}
Here we have used the identity $(dF)_{abc}=0$ and the
4-dimensional curvature 2-forms
\begin{displaymath}
\hat{\Omega}_{ab}^{\ \ ij}= (d\hat{\omega}^{ij})_{ab}+
\hat{\omega}^{i}_{\ k a}\ \wedge\ \hat{\omega}^{k j}_{\ \
b}\equiv\frac12\ \hat{\Omega}^{\ \ ij}_{kl} \
(\hat{e}^{k})_a\wedge\ (\hat{e}^l)_b.
\end{displaymath}

These geometrical considerations become physically relevant when
we postulate that gravitation in the five-dimensional space is
governed by the corresponding Palatini action (\ref{action}).
Using Eqs. (\ref{wabi4}) and (\ref{wabij}), the action
(\ref{action}) on $M$ can be reduced to the following action on
$S$:
\begin{eqnarray}
&&S_p[(\hat{e}_i)^a,\,\hat{\omega}^{ij}_{\ \
a},\,\lambda,\,F_{ab}]=
\nonumber\\
&& \int_S\ \lambda^{\frac12}\,\hat{e}
[(\hat{e}_i)^a(\hat{e}_j)^b\hat{\Omega}_{ab}^{\ \
ij}-\frac14\lambda F_{ab}F^{ab}-
2\lambda^{-\frac12}(\hat{e}_{i})^a\tilde{D}_aD^{i}(\lambda^{\frac12})],\label{Sp}
\end{eqnarray}where $\hat{e}$ is the square root of the determinant of $h_{ab}$.
Note that
\\ $(\hat{e}_{i})^a\tilde{D}_aD^{i}(\lambda^{\frac12})\neq
\tilde{D}_aD^{a}(\lambda^{\frac12})$. Now we define
\begin{displaymath}
\hat{C}_{ai}^{\ \ j}\equiv h^b_{a}C_{bi}^{\ \ j},
\end{displaymath}
where $C_{ai}^{\ \ j}$ comes from $C_{a\mu}^{\ \ \nu}$ defined by
Eq.(\ref{C1}). It is then easy to see
\begin{displaymath}
(\tilde{D}_{a}-D_{a})V_{i}=\hat{C}_{ai}^{\ \ j}V_{j}.
\end{displaymath}
Thus one has
\begin{equation}
\hat{\Omega}_{ab}^{\ \ ij}-\hat{R}_{ab}^{\ \ ij}
=2D_{[a}\hat{C}_{b]}^{\ ij}+2\hat{C}_{[a}^{\ ik}\hat{C}_{b]k}^{\ \
\ j},
\end{equation}
where $\hat{R}_{ab}^{\ \ ij}$ denotes internal curvature of $D_a$.

 Then the action (\ref{Sp}) becomes
\begin{eqnarray}
&&S_p[(\hat{e}_i)^a,\,\hat{C}^{\ \ j}_{ai},\,\lambda,\,F_{ab}]=
\nonumber\\
&& \int_S\
\hat{e}\Big[\lambda^{\frac12}\,\big((\hat{e}_i)^a(\hat{e}_j)^b\,
(\hat{R}_{ab}^{\ \ ij}\,+\,2D_{[a}\hat{C}_{b]}^{\ \ ij}\,+\,
2\hat{C}_{[a}^{\ ik}\,\hat{C}_{b]k}^{\ \ j})
\nonumber\\
&&-\frac14\lambda\,F_{ab}F^{ab}\big)-
2D^2(\lambda^{\frac12})+\,2(\hat{e}_i)^a \hat{C}_{aj}^{\ \
i}D^j(\lambda^{\frac12})\Big],\label{action1}
\end{eqnarray}
where $D^2=D_aD^a$ is the four-dimensional d'Alembertian operator.

 The sum of the second and the sixth terms reads:
\begin{equation}
2\lambda^{\frac12}(\hat{e}_i)^a(\hat{e}_j)^bD_{[a}\hat{C}_{b]}^{\
\ ij}+2(\hat{e}_i)^a \hat{C}_{aj}^{\ \ i}D^j(\lambda^{\frac12}) =
2D_a(\lambda^{\frac12}\hat{C}_b^{\
ij})(\hat{e}_i)^a(\hat{e}_j)^b.\label{sum}
\end{equation}

Since $D_{a}$ annihilates the tetrad, Eq. (\ref{sum}) and the
fifth term are pure divergences and therefore do not contribute to
the variation. Neglecting the boundary terms, action
(\ref{action1}) becomes:
\begin{eqnarray}
&&S_p[(\hat{e}_i)^a,\,\hat{C}^{\ \ j}_{ai},\,\lambda,\,F_{ab}]=
\nonumber\\
&& \int_S\
\hat{e}\lambda^{\frac12}\Big[(\hat{e}_i)^a(\hat{e}_j)^b\,
(\hat{R}_{ab}^{\ \ ij}\,+\, 2\hat{C}_{[a}^{\ ik}\,\hat{C}_{b]k}^{\
\ j})-\frac14\lambda\,F_{ab}F^{ab}\Big],\label{action2}
\end{eqnarray}
The variation of this action with respect to $\hat{C}_{ai}^{\ \
j}$ yields:
\begin{displaymath}
\big((\hat{e}_{k})^{[a}(\hat{e}_{l})^{b]} \delta^{k}_{[i}
\delta_{j]}^{m}\big)\,\hat{C}_{bm}^{\ \ l}=0 ,
\end{displaymath} which has the same form as Eq. (\ref{last}), implying  that:
\begin{equation}
\hat{C}_{ai}^{\ \ j}\,=\,0\label{4C0}.
\end{equation}

Hence the equation of motion for the connection $\tilde{D}_a$ is
again that it equals $D_a$. It is then straightforward to see that
action (\ref{action2}) gives exactly the same fields equations for
the dynamical variables $(\hat{e}_i)^a$, $\lambda$ and $A_a$ as in
5-dimensional Kaluza-Klein theory\cite{kk}. One can also
substitute Eq. (\ref{4C0}) for action (\ref{action2}) and get the
conventional reduced Kaluza-Klein action:
\begin{equation}
S_p[(\hat{e}_i)^a,\,\lambda,\,F_{ab}]= \int_S\
\lambda^{\frac12}\,\hat{e}[ (\hat{e}_{i})^a\,(\hat{e}_{j})^b\,
\hat{R}_{ab}^{\ \ ij} \,-\,\frac14\lambda\,F_{ab}F^{ab}].
\end{equation}

In conclusion, we have studied the reduction of the Palatini
action in 5-dimensional spacetime with a spacelike Killing vector
field. The 4-dimensional electromagnetic and scalar fields are
assumed to be related to the 5-dimensional Lorentz connection by
Eqs. (\ref{w4[ij]}) and (\ref{w4i4}). The reduced action
(\ref{Sp}) is in the 4-dimensional Palatini formalism of gravity
coupled to the electromagnetic field and the scalar field. It
gives the same equations of motion of Kaluza-Klein theory. The
reduced Palatini action for 5-dimensional Kaluza-Klein theory is
thus obtained. The reduction scheme might also be extended to the
Palatini formalism of higher-dimensional Kaluza-Klein theories.
Thus one may study Kaluza-Klein theory in Palatini formalism as
well, where the scaler field might play an important role in the
explanation to the dark physics in cosmology and the Higgs field
in particle physics.

\section*{ Acknowledgments}

This work is  supported in part by NSFC (10205002) and YSRF for
ROCS, SEM. You Ding and Muxin Han would also like to acknowledge
support from Undergraduate Research Foundation of BNU.

\end{document}